\documentclass[aps,prd,onecolumn,superscriptaddress,nofootinbib,floatfix,preprintnumbers]{revtex4-2}

\usepackage{url,hyperref,microtype,subcaption,booktabs,array}
\usepackage{amsmath,amssymb,amsfonts}
\usepackage{graphicx}

\begin{document}

\title{Thermofractal Dynamics in the Multisource Thermal Model: A Unified Approach to Multiparticle Production}

\author{Airton Deppman}
\email{deppman@if.usp.br}
\affiliation{Instituto de Física, Universidade de São Paulo, Rua do Matão 1371, CEP 05508-090, São Paulo, SP, Brazil}
\affiliation{Institute of Physics, Jan Kochanowski University, ul. Uniwersytecka 7, 25-406 Kielce, Poland}

\author{Trambak Bhattacharyya}
\email{trambak.bhattacharyya@ujk.edu.pl}
\affiliation{Institute of Physics, Jan Kochanowski University, ul. Uniwersytecka 7, 25-406 Kielce, Poland}

\author{Maciej Rybczyński}
\email{maciej.rybczynski@ujk.edu.pl}
\affiliation{Institute of Physics, Jan Kochanowski University, ul. Uniwersytecka 7, 25-406 Kielce, Poland}

\author{Grzegorz Wilk}
\email{Grzegorz.Wilk@ncbj.gov.pl}
\affiliation{National Centre for Nuclear Research, 00-681, Warsaw, Poland}

\author{Zbigniew W\l odarczyk}
\email{zbigniew.wlodarczyk@ujk.edu.pl}
\affiliation{Institute of Physics, Jan Kochanowski University, ul. Uniwersytecka 7, 25-406 Kielce, Poland}

\date{\today}

\begin{abstract}
The empirical success of non-extensive Tsallis statistics and multisource thermal models in describing transverse-momentum ($p_T$) spectra and rapidity distributions in high-energy relativistic collisions highlights a deep connection between expansion geometry, non-extensive thermodynamics, and QCD dynamics. In this work, we establish a theoretical framework showing that the thermofractal structure of QCD phase space provides a direct microscopic foundation for the multisource thermal model. By exploiting the Euler integral representation of non-extensive statistical factors, we demonstrate that a single thermofractal emission source generating $q$-exponential distributions is mathematically equivalent to a continuous superposition of local Boltzmann thermal sources whose temperature fluctuations follow an inverse-Gamma distribution. Furthermore, we prove that multi-component phenomenological fits with discrete source temperatures represent the exact, optimal Gauss-Laguerre quadrature discretisation of this underlying continuous thermofractal field. We show how this continuous field constrains the temperature parameter space in multisource formulations, matches pQCD hard-scattering power-law asymptotics, and provides a robust vacuum baseline for medium-induced jet quenching and nuclear modification factor ($R_{AA}$) calculations.
\end{abstract}

\maketitle

\section{Introduction}

Describing multiparticle production in high-energy proton-proton (\(pp\)), proton-nucleus (\(pA\)), and nucleus-nucleus (\(AA\)) collisions remains a central challenge in relativistic heavy-ion physics \cite{Liu2013,Liu2012,Deppman2016}. Over decades of experimental exploration at the Relativistic Heavy Ion Collider (RHIC) and the Large Hadron Collider (LHC), two distinct yet successful phenomenological paradigms have emerged to address different aspects of final-state hadron spectra. On one hand, the Multisource Thermal Models account for geometric, flow, and longitudinal expansion effects in relativistic heavy-ion and hadronic interactions. These models treat the emission volume as a spatio-temporal superposition of multiple thermal fireballs, participant partons, or expanding cylindrical sources \cite{Liu2013, Liu2014}. While visually intuitive and effective at capturing transverse expansion and longitudinal rapidity profiles, standard multisource formulations rely on tuning several independent source temperatures (\(T_1, T_2, T_3, \dots\)) and relative normalisation weights to reproduce low-, intermediate-, and high-\(p_T\) regions \cite{Liu2013}.

On the other hand, the Thermofractals and Tsallis Statistics, grounded in the non-extensive entropy proposed by Tsallis \cite{Tsallis1988}, the thermofractal framework \cite{Deppman2016, Deppman2018_Entropy, Deppman2020} provides a thermodynamic mechanism for \(q\)-exponential energy distributions based on scale invariance. Systematic empirical analyses across RHIC and LHC energies \cite{Azmi2014, Marques2013, Biro2017, Rocha2022} have demonstrated that energy fluctuations in self-similar QCD phase space naturally generate $q$-exponential distributions in \(p_T\) \cite{Deppman2016, Azmi2014}. Significantly, thermofractal theory predicts an entropic index \(q = 8/7 \approx 1.143\) derived from \(SU(3)\) phase-space effective degrees of freedom \cite{Deppman2020}. The incorporation of \(q\)-statistics and thermofractal concepts in expanding geometry is grounded in fundamental non-equilibrium dynamics \cite{Deppman2016, Wilk2000, Wilk2009, Rybczynski2020, Rybczynski2026, Beck2003}, and studies on local temperature fluctuations demonstrated that non-extensive statistics emerge naturally from superstatistical thermal ensembles \cite{Wilk2000, Wilk2009, Rybczynski2020}, operating in close connection with Beck-Cohen superstatistics \cite{Beck2003}. Modern kinetic theory shows that non-extensive statistics also arise from non-local correlations in quantum field theory and non-linear transport dynamics governed by the Plastino--Plastino equation \cite{Plastino1995}, which has been formally derived from the Boltzmann transport equation endowed with $q$-deformed algebraic correlations in self-similar thermofractal systems \cite{DeppmanPLB2023,Bhattacharyya2023,Megas2024}.

In this paper, we show that a thermofractal source governed by \(q\) is mathematically equivalent to a continuous superposition of local Boltzmann sources with inverse-Gamma distributed temperatures, establishing a theoretical connection to the multisource thermal model.

\section{Thermofractals and Multisource Models}

\subsection{Thermofractals and Tsallis Non-Extensive Statistics}

Standard Boltzmann-Gibbs thermodynamics relies on additive entropy, \(S = -k_B \sum_i p_i \ln p_i\), suitable for short-range interactions and uncorrelated phase-space volumes. In high-energy hadronic collisions, long-range color force correlations, scale invariance, and self-similar parton branching violate statistical independence \cite{Tsallis1988, Deppman2016}. Tsallis non-extensive statistics generalizes the entropy through a non-extensive entropic index \(q\):
\begin{equation}
S_q = k_B \, \frac{1 - \sum_{i} p_i^q}{q - 1},
\label{eq:tsallis_entropy}
\end{equation}
which recovers standard Boltzmann-Gibbs entropy in the limit \(q \to 1\).

Thermofractals represent a physical realization of non-extensive thermodynamics within quantum chromodynamics \cite{Deppman2016, Deppman2018_Entropy}. A thermofractal is defined as a thermodynamic system exhibiting self-similarity and scale invariance across its phase space \cite{Deppman2016}. When high energy is deposited into a collision volume, the system forms excited compound states (fireballs or partonic clusters) that recursively decay into smaller, self-similar sub-systems down through sequential generations \cite{Deppman2016, Deppman2020}.

The total internal energy \(U\) of a thermofractal system is partitioned between the kinetic energy \(K\) of \(N_{\text{sub}}\) constituent sub-systems and internal energy \(\varepsilon_i\) stored within internal compound states \cite{Deppman2016}:
\begin{equation}
U = K + \sum_{i=1}^{N_{\text{sub}}} \varepsilon_i,
\end{equation}
where \(K = \sum_{i=1}^{N_{\text{sub}}} K_i\). The kinetic state density is given by \(\rho(K) = A_0 K^{\alpha - 1}\) with \(\alpha = N_{\text{sub}} \cdot d\), where \(d\) denotes the spatial dimensionality of the kinetic phase space \cite{Deppman2016}. The scale-invariant phase-space density \(\Omega(U)\) obeys the recursive convolution integral \cite{Deppman2016, Deppman2018_Entropy}:
\begin{equation}
\Omega(U) = \int_{0}^{U} dK \, \rho(K) \int \left( \prod_{i=1}^{N_{\text{sub}}} d\varepsilon_i \, \Omega(\varepsilon_i) \right) \delta\left( U - K - \sum_{i=1}^{N_{\text{sub}}} \varepsilon_i \right).
\label{eq:convolution}
\end{equation}

The internal density of states scales as a power law governed by an effective phase-space parameter \(D\): \(\Omega(\varepsilon_i) \propto \varepsilon_i^{D-1}\) \cite{Deppman2016, Deppman2020}. In the thermodynamic limit, this power-law phase-space density yields the non-extensive energy distribution \(P(E)\) \cite{Deppman2016}:
\begin{equation}
\frac{d^2N}{dp_T dy} = C_q  ~p_T  ~m_T \cosh y  \left[ 1 + (q-1) \frac{m_T \cosh y - \mu}{T_0} \right]^{-\frac{q}{q-1}},
\label{eq:tsallis_energy_yield}
\end{equation}
where $C_q$ is the global normalisation factor, \(T_0\) is the scale temperature of the primary system, and \(q\) is linked to \(D\) by \cite{Deppman2016, Deppman2020}:
\begin{equation}
\frac{1}{q - 1} = D + 1 \implies q = 1 + \frac{1}{D + 1}.
\label{eq:q_fractal}
\end{equation}



Non-extensive statistical mechanics in expanding hadronic systems can also be formulated via non-linear transport dynamics \cite{Plastino1995}. The distribution function \(P(E, t)\) evolves according to the generalized Plastino--Plastino Fokker--Planck equation \cite{Plastino1995}:
\begin{equation}
\frac{\partial f(p, t)}{\partial t} = -\frac{\partial}{\partial p_i} \left[ A({\mathbf{p}}) f(p,t) \right] + \frac{\partial^2}{\partial p_i \partial p_j} \left[ B(\mathbf{p}) f^{2-q}(p, t) \right],
\label{eq:plastino_equation}
\end{equation}
In momentum space, the distribution \(P(p,t)\) evolves according to the generalised Plastino–Plastino Fokker–Planck equation, with linear drift \(A(p)=-\gamma p\) and constant diffusion coefficient \(B(\mathbf{p})=B\). The stationary solution depends on the drift and diffusion coefficients and on the boundary conditions. Its relationship to the thermofractal energy distribution in Eq. (4) must therefore be established separately. The stationary solution of Eq.~(\ref{eq:plastino_equation}) yields \(q\)-exponential energy distributions \cite{Plastino1995, Wilk2000}, and has the form given in Eq~(\ref{eq:tsallis_energy_yield})\footnote{The actual form of the stationary distribution used to fit data may vary according to the definition of $q$.}. Nonextensive dynamics has also been used to study rapidity distributions in high-energy collisions~\cite {Bhattacharyya2026}. 

\subsection{The Multi-Source Thermal Model: Physical and Geometric Foundations}

The multi-source thermal model provides a macroscopic hydrodynamic and geometric framework to describe multiparticle momentum and rapidity distributions in high-energy hadronic (\(pp\)) and relativistic heavy-ion (\(AA\)) collisions~\cite{Liu2013,Liu2012}. Rather than assuming that the entire interaction volume forms a single static fireball in global equilibrium, the multi-source model treats the emitting system as an inhomogeneous, spatio-temporal superposition of multiple localised emission sub-sources~\cite{Liu2013,Liu2012}.


In the multi-source model, the total single-particle differential yield across transverse momentum \(p_T\) and rapidity \(y\) is given by a weighted linear combination of \(N_{\text{src}}\) discrete thermal sources:
\begin{equation}
\left(\frac{d^2N}{dp_T dy}\right)_{\text{MS}} = \sum_{i=1}^{N_{\text{src}}} k_i \mathcal{S}_i(p_T, y; T_i, \beta_T, \mu),
\end{equation}
where \(k_i\) is the relative yield contribution of the \(i\)-th source (\(\sum_{i=1}^{N_{\text{src}}} k_i = 1\))~\cite{Liu2013}.

For a single thermal source at local temperature $T_i$ and chemical potential $\mu$, particle emission follows relativistic phase-space kinetics. The local invariant thermal spectrum $\mathcal{S}_i$ takes the form\cite{Liu2012}:
\begin{equation}
\mathcal{S}i(p_T, y; T_i, \mu) = C(T_i) \, p_T m_T \cosh(y - y_{b,i}) \exp\left(-\frac{m_T \cosh(y - y_{b,i}) - \mu}{T_i}\right),
\label{eq:thermal_source_norm}
\end{equation}
where $C(T_i)$ is the temperature-dependent normalization factor. When superimposing thermal sources, distinction must be made between combining unnormalized Boltzmann exponential factors—which directly maps onto the Euler integral representation—and combining fully normalized thermal yields $\mathcal{S}_i$, where $C(T_i)$ introduces additional temperature weighting into the superposition kernel.

When collective transverse radial expansion with velocity \(\beta_T\) (\(\gamma_T = 1/\sqrt{1 - \beta_T^2}\)) is present, the effective single-particle energy in the rest frame of the expanding fluid cell is Lorentz-boosted:
\begin{equation}
E_{s,i} = \gamma_T \left[ m_T \cosh(y - y_{b,i}) - p_T \beta_T \right] - \mu.
\end{equation}
The longitudinal extension of source rapidities \(y_b\) is parameterized by a Gaussian profile centered at mid-rapidity \(y_0\) with width \(\sigma_y\):
\begin{equation}
f_y(y_b) = \frac{1}{\sqrt{2\pi}\sigma_y} \exp\left( -\frac{(y_b - y_0)^2}{2\sigma_y^2} \right).
\end{equation}

Although multi-temperature parameterizations (\(N_{\text{src}} = 2\) or $3$, with \(T_1 \approx 100\text{--}120~\text{MeV}\), \(T_2 \approx 180\text{--}220~\text{MeV}\), and \(T_3 \ge 300~\text{MeV}\)) fit transverse momentum spectra across soft and hard regimes, the conventional multi-source model treats \(T_i\) and \(k_i\) as independent free parameters, lacking a microscopic thermodynamic principle that governs their discrete values~\cite{Liu2013}.

\section{Decomposition of Thermofractals into Thermal Superpositions}

Using Euler's integral identity for the Gamma function,
\begin{equation}
x^{-a} = \frac{1}{\Gamma(a)} \int_{0}^{\infty} t^{a-1} e^{-x t} \, dt \quad (x > 0, \, a > 0),
\label{eq:gamma_identity}
\end{equation}
we set \(x = 1 + (q-1)\dfrac{E}{T_0}\) and \(a = \dfrac{q}{q-1}\), where \(E = m_T \cosh y - \mu\). Substituting these definitions into the equation above yields:
\begin{equation}
\left[ 1 + (q-1)\frac{E}{T_0} \right]^{-\frac{q}{q-1}} = \frac{1}{\Gamma\left(\frac{q}{q-1}\right)} \int_{0}^{\infty} t^{\frac{q}{q-1}-1} \exp\left( -t \left[ 1 + (q-1)\frac{E}{T_0} \right] \right) dt.
\label{eq:integral_step1_gen}
\end{equation}

Defining the fluctuating local temperature variable \(T_s \equiv \frac{T_0}{(q-1) t}\), we transform the integration variable from \(t\) to \(T_s\):
\begin{equation}
\frac{d^2N}{dp_T dy} = C_q ~p_T \, m_T \int_{0}^{\infty} f(T_s; T_0, q) \, \exp\left( -\frac{m_T \cosh y - \mu}{T_s} \right) dT_s,
\label{eq:superposition_identity_gen}
\end{equation}
where \(f(T_s; T_0, q)\) is the normalized inverse-Gamma probability density function \cite{Wilk2000, Beck2003}:
\begin{equation}
f(T_s; T_0, q) = \frac{1}{\Gamma\left(\frac{q}{q-1}\right)} \left( \frac{T_0}{q-1} \right)^{\frac{q}{q-1}} \frac{1}{T_s^{\frac{q}{q-1} + 1}} \exp\left( -\frac{T_0}{(q-1) T_s} \right).
\label{eq:inverse_gamma_gen}
\end{equation}

Eq. (\ref{eq:superposition_identity_gen}) demonstrates that a $q$-exponential thermofractal factor maps into a continuous superposition of local Boltzmann exponentials weighted by an inverse-Gamma distribution $f(T_s; T_0, q)$\cite{Wilk2000, Beck2003, Rybczynski2020}. Note that if local Boltzmann sources are normalised individually prior to superposition, the temperature-dependent normalisation factor $C(T_s)$ is absorbed into an effective weight function $\tilde{f}(T_s) \propto C(T_s) f(T_s; T_0, q)$, shifting the effective quadrature nodes accordingly.

Defining the inverse temperature variable \(\beta \equiv 1/T_s\), the distribution \(f(T_s; T_0, q) dT_s\) converts into a Gamma distribution \(g(\beta) d\beta\):
\begin{equation}
g(\beta) \, d\beta = \frac{1}{\Gamma(n)} \left( \frac{n}{\beta_0} \right)^n \beta^{n-1} e^{-\frac{n \beta}{\beta_0}} d\beta, \quad \text{where } n = \frac{q}{q-1} \text{ and } \beta_0 = \frac{q}{T_0}.
\label{eq:gamma_beta}
\end{equation}

The expectation value of \(\beta\) is \(\langle \beta \rangle = \beta_0 = q/T_0\). Within Beck-Cohen Superstatistics \cite{Beck2003, Wilk2000}, the relative variance of inverse-temperature fluctuations across local sources is given by:
\begin{equation}
\frac{\langle \beta^2 \rangle - \langle \beta \rangle^2}{\langle \beta \rangle^2} = \frac{1}{n} = \frac{q - 1}{q}.
\label{eq:variance_q}
\end{equation}
For \(q = 8/7\), the relative fluctuation variance is fixed to \((q-1)/q = 1/8 = 12.5\%\).

To evaluate or discretise the exact continuous temperature field given by Eq. (14) for practical applications in multisource model fits, one can apply a discrete quadrature approximation. This method provides an exact transition from the continuous thermofractal temperature field \(f(T_s; T_0, q)\) to discrete \(N_{\text{src}}\)-component multisource models.

Substituting \(x = \frac{T_0}{(q-1) T_s}\) into Eq.~(\ref{eq:superposition_identity_gen}) maps the thermal superposition integral directly onto the standard Gauss-Laguerre weight function:
\begin{equation}
\frac{d^2N}{dp_T dy} = C_q ~\frac{p_T \, m_T}{\Gamma(\alpha+1)} \int_{0}^{\infty} x^{\alpha} e^{-x} \exp\left( -x \frac{(q-1)E}{T_0} \right) dx,
\label{eq:gauss_laguerre_integral}
\end{equation}
where the shape exponent is uniquely fixed by the entropic index:
\begin{equation}
\alpha = \frac{q}{q-1} - 1 = \frac{1}{q-1}.
\end{equation}

For \(q = 8/7\), the exponent evaluates to \(\alpha = 7\). The integral in Eq.~(\ref{eq:gauss_laguerre_integral}) can be evaluated numerically with high precision using an \(N_{\text{src}}\)-point Gauss-Laguerre quadrature:
\begin{equation}
\int_{0}^{\infty} x^{\alpha} e^{-x} H(x) \, dx \approx \sum_{i=1}^{N_{\text{src}}} w_i \, H(x_i),
\label{eq:quadrature_sum}
\end{equation}
where \(x_i\) are the roots of the generalized Laguerre polynomial \(L_{N_{\text{src}}}^{(\alpha)}(x)\), and \(w_i\) are the corresponding quadrature weights. The discrete source temperatures \(T_i\) extracted in phenomenological multisource fits thus directly correspond to the optimal quadrature nodes:
\begin{equation}
T_i = \frac{T_0}{(q-1) x_i}, \quad \text{with relative weights } k_i = \frac{w_i}{\Gamma(\alpha+1)}.
\label{eq:quadrature_nodes}
\end{equation}

This mapping highlights the relationship between continuous thermofractal dynamics and discrete multisource models. While the representation of the $q$-exponential Tsallis factor as a continuous superposition of Boltzmann exponentials over a Gamma inverse-temperature distribution in Eq.(~\ref{eq:gamma_beta})  is exact, the discrete $N_{\text{src}}$-component formulation obtained via Gauss-Laguerre quadrature in Eq.(~\ref{eq:quadrature_sum}) is a finite-degree numerical approximation. For small numbers of sources ($N_{\text{src}} = 2$ or $3$), this quadrature provides an optimal discrete approximation to the continuous thermofractal field, rather than an exact algebraic identity.

Up to this point, the thermofractal superposition integral has been formulated in the rest frame of a single, static emission source. In relativistic hadronic and heavy-ion collisions, however, the continuous thermofractal temperature field $f(T_s; T_0, q)$ is embedded within a hydrodynamically expanding medium characterised by transverse radial flow and longitudinal expansion along the beam axis.

To account for collective radial expansion with transverse velocity $\beta_T$ (Lorentz factor $\gamma_T = 1/\sqrt{1 - \beta_T^2}$), assuming flow aligned with $p_T$, the effective energy of an emitted particle in the rest frame of a local fluid cell is modified to $E_s = \gamma_T \left[ m_T \cosh(y - y_b) - p_T \beta_T \right] - \mu$, where $y_b$ denotes the longitudinal shift rapidity of the emitting cell. Convolving the single-cell emission yield over both the longitudinal source rapidity distribution $f_y(y_b)$ and the continuous thermofractal temperature distribution $f(T_s; T_0, q)$ yields the master differential spectrum:
\begin{equation}
\frac{d^2N}{dp_T dy} = \int_{-\infty}^{\infty} dy_b \int_{0}^{\infty} dT_s \; f_y(y_b) \, f(T_s; T_0, q) \, C_q ~p_T ~m_T \exp\left( -\frac{\gamma_T \left[ m_T \cosh(y - y_b) - p_T \beta_T \right] - \mu}{T_s} \right).
\label{eq:master_equation}
\end{equation}

For a longitudinally expanding system described by $N_{\text{cyl}}$ cylindrical sources centered at rapidities $y_{b,j}$ with spatial widths $\sigma_{y,j}$ \cite{Liu2013, Liu2014}:
\begin{equation}
f_y(y_b) = \sum_{j=1}^{N_{\text{cyl}}} \frac{k_j}{\sqrt{2\pi}\sigma_{y,j}} \exp\left( -\frac{(y_b - y_{b,j})^2}{2\sigma_{y,j}^2} \right).
\label{eq:rapidity_multicylinder}
\end{equation}
In the limit of vanishing radial flow ($\beta_T \to 0, \gamma_T \to 1$) and localized rest-frame rapidity emission $f_y(y_b) \to \delta(y_b)$, Eq.~(\ref{eq:master_equation}) reduces smoothly to the single-source Lorentz-invariant spectrum. 

\begin{figure}[h!]
\centering
\includegraphics[width=0.48\textwidth]{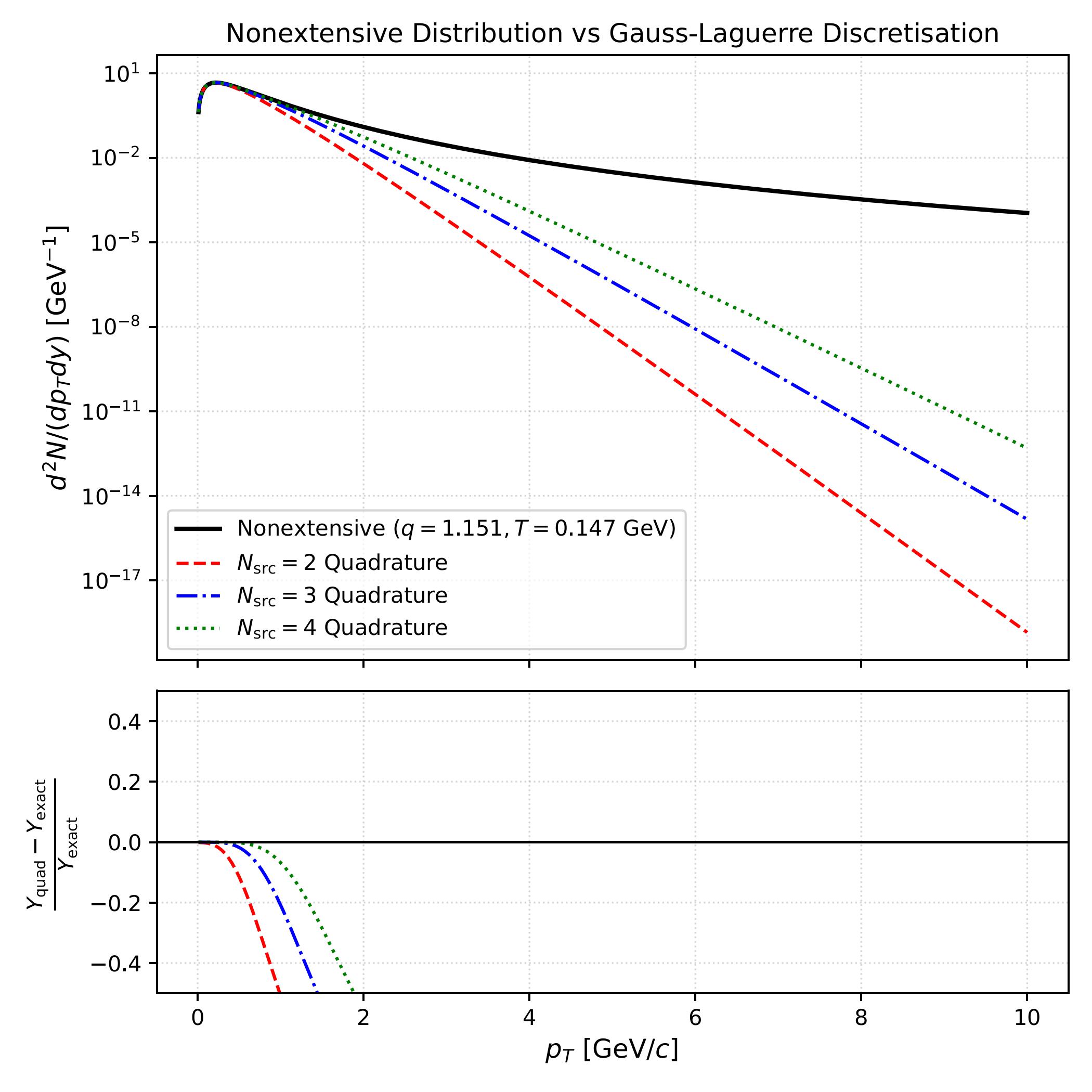}
\caption{Comparison of the exact continuous nonextensive yield~(Ref.~\cite{Wong2015} with $N_{\text{src}} = 2, 3,$ and $4$ Gauss-Laguerre quadrature approximations for $q = 8/7$. The lower panel shows the relative deviations as a function of $p_T$.}
\label{fig:quadrature_test}
\end{figure}

To assess the accuracy and range of validity of the discrete multisource representation, we compare, in Figure~\ref{fig:quadrature_test}, the exact continuous thermofractal yield [Eq.~(\ref{eq:superposition_identity_gen})] with $N_{\text{src}}$-component Gauss-Laguerre quadrature approximations [Eq.~(\ref{eq:quadrature_sum})] for $q = 8/7$ ($\alpha = 7$). We emphasise that the exact curve representing the nonextensive distribution provides a good description of the experimental data over 14 orders of magnitude of the measured cross-section, covering the $p_T$ range $0<p_T<1000$~GeV~\cite{Wong2015}. The multi-source, as well, model has been tested in several works (see~\cite{Olimov2022,Chen2023} and references therein).

For $N_{\text{src}} = 2$, the roots of $L_2^{(7)}(x)$ yield discrete temperatures $T_1 = 1.1667\,T_0$ and $T_2 = 0.5833\,T_0$ with relative normalization weights $k_1 = 2/3$ and $k_2 = 1/3$. In Figure~\ref{fig:quadrature_test}, we show the invariant differential yield $d^2N/(dp_T dy)$ as a function of $p_T$ for $N_{\text{src}} = 2, 3,$ and $4$ compared against the exact continuous thermofractal result, alongside the corresponding relative deviations $\left(Y_{\text{quad}} - Y_{\text{exact}}\right) / Y_{\text{exact}}$.

The comparison demonstrates that while a two-source formulation captures the low- and intermediate-$p_T$ spectrum within $5\%$ accuracy up to $p_T / T_0 \approx 10$, increasing to $N_{\text{src}} = 3$ or $4$ extends the agreement to higher $p_T$ scales. At ultra-high $p_T$, any finite quadrature eventually underestimates the exact power-law tail, reflecting the physical necessity of the full continuous temperature field $f(T_s; T_0, q)$ for high-$p_T$ hard scattering.

\section{Discussion and Conclusions}

We have established a formal equivalence connecting thermofractal dynamics, Superstatistics, and the multisource thermal model. By demonstrating that \(q\)-exponential distributions map into a continuous superposition of Boltzmann sources through an inverse-Gamma temperature density \(f(T_s; T_0, q)\), we link microscopic phase-space self-similarity \cite{Deppman2016, Deppman2020} with macroscopic multi-source expansion geometries \cite{Liu2013, Liu2014}.

Key insights obtained from this synthesis include:
\begin{enumerate}
\item \textbf{Parameter Constraint:} Fixing \(q = 8/7 \approx 1.143\) from \(SU(3)\) thermofractal theory constrains the temperature distribution shape \(f(T_s)\), replacing unconstrained multi-temperature parameters (\(T_1, T_2, T_3, \dots\)) with a single primary scale parameter \(T_0\).

\item \textbf{Fluctuation Variance:} The relative variance of inverse-temperature fluctuations across local sources is fixed to \((q-1)/q = 1/8 = 12.5\%\).
\item \textbf{High-$p_T$ Asymptotics and pQCD Connection:} In the ultrarelativistic regime ($p_T \gg m_0$), the single-particle differential yield scales asymptotically as:
\begin{equation}
\frac{d^2N}{dp_T dy} \propto p_T ~m_T \cdot p_T^{-\frac{q}{q-1}} = p_T^2 \cdot p_T^{-8} = p_T^{-6}.
\label{eq:asymptotic_power_law}
\end{equation}
For $q = 8/7$, this power-law falloff ($p_T^{-6}$) is in direct agreement with the dimensional counting rules and perturbative QCD (pQCD) parton $2 \to 2$ hard-scattering expectations observed across RHIC and LHC $pp$ spectra. This demonstrates that thermofractal dynamics smoothly bridges soft thermal particle production at low $p_T$ and hard partonic scattering tails at high $p_T$ without requiring ad hoc model switches. Furthermore, while finite Gauss-Laguerre quadratures ($N_{\text{src}} = 2, 3$) serve as effective discrete approximations capturing the low- and intermediate-$p_T$ regions, they truncate the spectrum at high momentum. Resolving the exact, continuous power-law asymptotics ($p_T \to \infty$) highlights the physical necessity of the full, continuous inverse-Gamma distribution $f(T_s; T_0, q)$.

\item \textbf{Medium Effects and Nuclear Modification Factor ($R_{AA}$):}
In heavy-ion collisions ($AA$), a dense partonic medium modifies particle production relative to vacuum $pp$ interactions. Within the thermofractal paradigm, medium effects do not simply wash out non-extensive statistics towards BG equilibrium; instead, partonic energy loss, medium-induced jet quenching, and non-local correlations alter the energy structure across subsystem generations. Recent transport analyses based on the Plastino--Plastino Fokker--Planck equation have demonstrated that the nuclear modification factor $R_{AA}(p_T)$ and energy-loss transport parameters (such as the jet quenching coefficient) can be rigorously calculated by modelling the anomalous diffusion of partons traversing the non-additive QGP \cite{BhattacharyyaPLB2024, BaptistaEPJA2025}. The continuous inverse-Gamma distribution $f(T_s; T_0, q)$ established in this work serves as the exact baseline vacuum field, upon which medium-drag interaction kernels modulate the effective energy scale $T_0 \to T_0^{\text{med}}(p_T)$, providing a unified physical connection between $pp$ vacuum spectra, transport dynamics in expanding fireballs, and $R_{AA}$ suppression patterns observed across RHIC and LHC energies.
\end{enumerate}

In summary, the unified thermofractal-multisource framework provides a consistent mathematical bridge connecting scale-invariant QCD phase-space dynamics, temperature-fluctuating Superstatistics, and multi-source hadronic expansion models.

\begin{acknowledgments}
A. D. thanks the warm hospitality found at the Jan Kochanowski University. This work was partially supported by the Conselho Nacional de Desenvolvimento Científico e Tecnológico (CNPq) and Fundação de Amparo à Pesquisa do Estado de São Paulo (FAPESP) and  by the project RID/SP/00015/2024/01.
\end{acknowledgments}

\end{document}